\documentclass[a4paper,12pt]{article} \usepackage[english]{babel}
\usepackage{fullpage} \usepackage{epsfig} \usepackage{amsfonts}
\usepackage{amssymb} \usepackage{amsmath} \usepackage{theorem}
\usepackage{eufrak}

\numberwithin{equation}{section}

\newtheorem{theorem}{Theorem}[section]
\newtheorem{corollary}[theorem]{Corollary}

\newtheorem{proposition}[theorem]{Proposition} {
\theorembodyfont{\normalfont}  } {
\theorembodyfont{\normalfont}  }
\makeatletter \title{Proof of Nishida's conjecture on anharmonic lattices } \author{Bob Rink\thanks{Mathematics
Department, Imperial College London, E-mail: {\tt b.rink@ic.ac.uk}. The author is supported by an EPSRC postdoctoral fellowship.}} \begin{document}  \hyphenation{boun-da-ry mo-no-dro-my sin-gu-la-ri-ty ma-ni-fold ma-ni-folds re-fe-rence se-ve-ral dia-go-na-lised con-ti-nuous thres-hold mo-ving ener-gy Birk-hoff si-mi-lar li-nea-ri-sa-tion un-ex-pec-ted-ly}
\maketitle
\abstract{\noindent We prove Nishida's 1971 conjecture stating that almost all low-energetic motions of the anharmonic Fermi-Pasta-Ulam lattice with fixed endpoints are quasi-periodic. 
The proof is based on the formal computations of Nishida, the KAM theorem, discrete symmetry considerations and an algebraic trick that considerably simplifies earlier results. }
\section{Introduction}
\begin{figure}[h]  \centering \includegraphics[width=10cm, height=2cm]{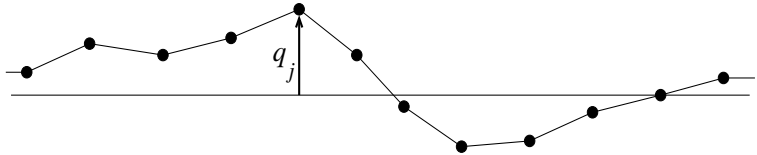}\renewcommand{\figurename}{\rm \bf \footnotesize Figure}
\caption{\footnotesize Schematic picture of the FPU lattice.}
\label{plaatjeketen}
\end{figure}

The Fermi-Pasta-Ulam (FPU) lattice is the famous discrete model for a continuous nonlinear string, introduced by E. Fermi, J. Pasta and S. Ulam \cite{Alamos}. It consists of a number of equal point masses that nonlinearly interact  with their  nearest neighbors. The physical variables of the FPU lattice are the positions $q_j$ of the particles, see Figure \ref{plaatjeketen}, and their conjugate momenta $p_j$. \\
 \vspace{-.3cm}

\noindent Fermi, Pasta and Ulam were interested in the statistical properties of the nonlinear FPU lattice. In fact, they expected that it would attain a thermal equilibrium, as was predicted by laws in statistical mechanics. This means that the initial energy of the lattice would be redistributed and, averaged over time, equipartitioned among all the Fourier modes of the lattice, see \cite{jackson}. They performed a numerical experiment to investigate how and at what time-scale this would occur. The astonishing result of their integrations was that there was no sign of
energy equipartition at all, see \cite{Alamos} and \cite{jackson}: energy that was initially put in one Fourier mode, was shared by only
a few other modes. Moreover, within a rather short time all the energy in the system returned
to the initial mode. This recurrent behaviour has been observed in experiments on the FPU lattice with quite small as well as very large numbers of particles, on short and long time-scales, and we are led to believe that at low energy the FPU lattice behaves more or less quasi-periodically. This observation was a big surprise. On the other hand, when the initial energy of the lattice is larger then a certain threshold, equipartition indeed occurs. \\
\indent  For a theoretical understanding of the Fermi-Pasta-Ulam experiment, one has always tried to link the FPU lattice to a completely integrable system. These are dynamical systems possessing a complete set of integrals of motion and therefore they display the regular type of behavior that was observed in the FPU experiment. With this in mind, two main approaches have been proposed.  Firstly, some unexpectedly regular phenomena were observed numerically by Zabuski and Kruskal \cite{Kruskal} in the Korteweg-de Vries (KdV) equation. It was later proved by Gardner et al. \cite{gardner} that the latter equation is integrable. In fact, Peter Lax realised that KdV is a member of a hierarchy of integrable equations that have a Lax-pair representation and therefore a complete set of integrals. See \cite{Palais} for a good overview of these results. Using the line of thought of Lax, Flashka \cite{Flaschka} could for instance prove the complete integrability of the so-called Toda-lattice, which is a finite-dimensional Hamiltonian system very similar to the FPU lattice. On the other hand, one may hope to construct an integrable {\it partial} differential equation describing the behavior of the FPU lattice with some accuracy. Thus one assumes the existence of a smooth interpolation function  $u: [0,1] \to \mathbb{R}$, and writes $q_j = u(j/n)$. If we now let $n$ grow to infinity and choose appropriate $n$-dependent scalings of the Hamiltonian function of the FPU system, we can -at least formally- obtain a partial differential equation for $u$ and hope to find that it is an integrable one. It is important to realise though that this procedure is not so well-defined: different ways of approximating the derivatives of $u$ may lead to different partial differential equations for $u$, see also \cite{jackson} in which among others a Boussinesq equation is obtained. Moreover, it is a priori not clear if the solutions of the resulting PDE constitute good approximations for $q$, even on finite time intervals. Although several heuristic arguments are available that link FPU to for instance KdV, see again \cite{jackson},  I have not been able to find any proof of such a statement.   \\
\indent The second approach differs greatly from that of the continuum approximations and is based on finite dimensional considerations -we shall pursue this approach in the rest of this paper. As is well-known \cite{arnold}, periodic and quasi-periodic motion is typical in completely integrable finite-dimensional Hamiltonian systems. Unfortunately, the finite dimensional FPU lattice does not belong to this category. One possible explanation of the recurrent behaviour of the lattice is therefore based on the famous 
Kolmogorov-Arnol'd-Moser (KAM) 
theorem \cite{arnold}, \cite{Broer}. This theorem explains   
that large measure Cantor sets of quasi-periodic motions can also exist in classes of nonintegrable Hamiltonian systems, namely small perturbations of certain integrable Hamiltonian systems. The only restrictive requirement is that the integrable system that we are perturbing satisfies a certain nondegeneracy condition, which requires that each quasi-periodic motion of the integrable system has a different frequency.  Even though various -again heuristic- arguments advocate this approach, and I mention in particular \cite{Chirikov}, the big problem is that it is not at all a priori clear whether 
the FPU lattice can really be viewed as a perturbation of such a {\it nondegenerate} integrable Hamiltonian system. The only obvious integrable approximation to the FPU lattice is its linearisation, which is highly degenerate as its frequency map is constant.  This problem was pointed out again recently in the excellent review paper by Ford \cite{Ford} and the book by Weissert \cite{Weissert},\\
\indent An interesting attempt to prove the applicability of the KAM theorem was made by T. Nishida \cite{Nishida}, who in 1971 published a paper that considers the FPU lattice with a finite number of particles, fixed endpoints and symmetric potential energy density function (the so-called $\beta$-lattice).  
Assuming a rather strong nonresonance condition on the frequencies of this lattice, Nishida computes its so-called Birkhoff normal form and shows that this normal form constitutes a nondegenerate integrable approximation to the original lattice Hamiltonian. Thus he proves the applicability of the KAM theorem and the existence of a positive measure set of quasi-periodic motions in the nonlinear FPU lattice. But all of this is under the assumption of a nonresonance condition, which unfortunately is only satisfied in exceptional cases. The actual value of Nishida's computation therefore remains unclear. \\ 
\indent This paper is nevertheless devoted to a full proof of what Nishida intended to show. The main result can be summarized as follows: \\ \mbox{} \\ 
\noindent {\it The Fermi-Pasta-Ulam lattice with fixed endpoints and an arbitrary finite number of moving particles possesses a completely integrable finite order Birkhoff normal form, which constitutes an integrable appoximation to the original Hamiltonian function. The integrals are the linear energies of the Fourier modes. When the potential energy density function of the lattice is an even function ($\beta$-lattice), this integrable approximation is nondegenerate in the sense of the KAM-theorem. This proves the existence of a large-measure set of quasi-periodic motions in the low-energy domain of the $\beta$-lattice.} \\ \mbox{} \\ \noindent The key to proving this result lies in the fact that Nishida's nonresonance condition, which a priori seems highly necessary for computing the Birkhoff normal form, is actually obsolete. As in \cite{Rink2}, \cite{Rink4}, which treat the FPU lattice with periodic boundary conditions, discrete symmetries are the key to proving Nishida's `conjecture'. The results of the present paper can be considered as an extension of \cite{Rink2} to the lattice with fixed endpoints with a considerably simpler proof which again uses discrete symmetry together with a simple algebraic trick. \\
\indent I want to remark here that the results of this paper do not provide any explicit bounds on the domain of validity of the normal form approximation. In particular we have at this moment no estimates on the behaviour of this domain when  $n$ grows to infinity. The principal interest of the result lies in the fact that, at least to my knowledge, it is the first complete proof of the very existence of quasi-periodic motion in the FPU lattice with fixed endpoints.

\section{The lattice equations of motion}
\indent As was mentioned before, the physical variables of the FPU lattice are the positions and conjugate momenta $(q,p)=(q_1, \ldots, q_n; p_1, \ldots, p_n)$ of the particles in the lattice, of which we assume here that there are only finitely many. These positions and momenta are elements of the $2n$-dimensional cotangent bundle $T^*\mathbb{R}^n \cong \mathbb{R}^{2n}$ of $\mathbb{R}^n$. $T^*\mathbb{R}^n$ is a symplectic manifold with the canonical symplectic form $dq\wedge dp:= \sum_{j=1}^ndq_j \wedge dp_j$. Given a Hamiltonian function $H: T^*\mathbb{R}^n \to \mathbb{R}$, the Hamiltonian vector field $X_H$ on $T^*\mathbb{R}^n$ is implicitly defined by the relation $dq \wedge dp (X_H, \cdot) =dH$. The integral curves of $X_H$ therefore are the solutions of the system of ordinary differential equations 
$$\dot q_j = \frac{\partial H}{\partial p_j} \ , \ \dot p_j = -\frac{\partial H}{\partial q_j} $$
In the case of the FPU lattice the Hamiltonian function is the sum of the kinetic energies of all the particles and the interparticle potential energies:
\begin{equation}\label{hamfpu} H = \sum_{j} \frac{1}{2} p_j^2 + W(q_{j+1} - q_j)  
\end{equation} in which $W: \mathbb{R} \to \mathbb{R}$ is traditionally a potential energy density function of
the form 
\begin{equation}\label{potentiaal}
W(x) = \frac{1}{2!} x^2 + \frac{\alpha}{3!} x^3 + \frac{\beta}{4!} x^4 
\end{equation}
The parameters $\alpha$ and $\beta$ measure the
nonlinearities in the forces between the particles in the lattice. \\
\indent The range of the summation index $j$ in (\ref{hamfpu}) depends on the exact boundary conditions that we impose on the lattice. For lattices with {\it fixed endpoints} the moving particles are labeled $j=1,\ldots, n$ and the endpoint particles are kept at rest, i.e.  $q_0 = q_{n+1} = 0$ for all time.  
When we impose  {\it periodic boundary conditions}, we label the particles by elements of the cyclic group, $j \in \mathbb{Z} / _{N \mathbb{Z}}$, so that the first and the last particle are identified, that is $q_0 = q_N$ for all time. For conveniency we have denoted the number of particles in a periodic lattice by $N$ (i.e. not by $n$), so that its phase space becomes $T^*\mathbb{R}^N$.
\section{Discrete symmetry}
The Hamiltonian function (\ref{hamfpu}) of the periodic FPU lattice (i.e. summation over $j \in \mathbb{Z} /_{N \mathbb{Z}}$) has discrete symmetries of which we shall discuss some dynamical consequences.  Two important symmetries of the periodic FPU lattice are the linear mappings $R, S:
T^*\mathbb{R}^{N} \to T^*\mathbb{R}^{N}$ defined by
\begin{align}\label{RSactie} \nonumber
R:&(q_1, q_2, \ldots, q_{N-1}, q_N;  p_1, p_2, \ldots, p_{N-1}, p_N)\! \mapsto \nonumber \\
&(q_{2}, q_3, \ldots, q_N, q_{1};  p_{2}, p_3, \ldots, p_N, p_1) \nonumber \\
S:&(q_1, q_2, \ldots, q_{N-1}, q_N;  p_1, p_2, \ldots, p_{N-1}, p_N)\! \mapsto \nonumber \\ 
&-(q_{N-1}, q_{N-2}, \ldots, q_1, q_{N};  p_{N-1},p_{N-2}, \ldots, p_1, p_N) \nonumber 
\end{align} 
It is easily checked that $R$ and $S$ are canonical transformations that leave the periodic FPU Hamiltonian (\ref{hamfpu}) invariant, i.e. $R^*(dq \wedge dp) = S^*(dq\wedge dp) = dq \wedge dp$ and $R^*H (=H \circ R) = S^*H (=H \circ S) = H$. This implies that $R^*X_H = X_{R^*H}= X_H$ and $S^*X_H = X_{S^*H} = X_H$, that is
$R$ and $S$ conjugate the Hamiltonian vector field $X_H$ to itself. This in turn implies that $R$ and $S$ commute with the time-$t$ flows $e^{tX_H}$ of $X_H$. Canonical diffeomorphisms with this property are called symmetries of $H$ and the group of symmetries of $H$ is denoted $G_H$. The subgroup $\langle R, S \rangle = \{{\rm Id}, R, R^2, \ldots, R^{N-1}, S, SR, SR^2, \ldots, SR^{N-1}\} \subset G_H$ is isomorphic to the $N$-th dihedral group, the symmetry group of the $N$-gon, as its elements satisfy the multiplication relations $R^N = S^2 = {\rm Id}$, $SR = R^{-1}S$. As $R$ and $S$ are linear mappings, the elements of $\langle R, S \rangle$ actually define a representation of $D_N$ in $T^*\mathbb{R}^N$ by symplectic mappings. \\
\indent  For every subgroup $G \subset G_H$, we define the fixed point set 
\begin{equation}
\mbox{Fix} \ \! G = \{ (q, p) \in T^*\mathbb{R}^N | P(q,p) = (q,p) \ \forall P \in G \} 
\end{equation}
Let $(q,p) \in {\rm Fix}\ \! G$ and $P \in G$. Then $P(e^{tX_H}(q,p)) = e^{tX_H}(P(q,p)) = e^{tX_H}(q,p)$, i.e. $e^{tX_H}(q,p) \in {\rm Fix}\ \! G$. Thus we see that ${\rm Fix} \ \! G$ is an invariant manifold for the flow of $X_H$.  
\begin{proposition} \label{compactinvariant}
When $G$ is compact and consists of linear symplectic isomorphisms of $T^*\mathbb{R}^n$, then ${\rm Fix} \ \! G$ is a symplectic manifold with the restriction to ${\rm Fix} \ \! G$ of $dq\wedge dp$ as symplectic form.  This implies that whenever $X_H$ is tangent to ${\rm Fix} \ \! G$, in particular when $H$ is $G$-symmetric, $$ (X_H)|_{{\rm Fix}\  \! G} = X_{(H|_{{\rm Fix}\ \!\!  G})} $$ 
\end{proposition}
{\bf Proof:} Clearly, ${\rm Fix} \ \! G = \cap_{P \in G} \ker (P - {\rm Id})$ is a subspace of $T^*\mathbb{R}^N$, and hence a submanifold. It remains to be proven that for every $(q,p) \in {\rm Fix} \ \! G$, the restriction of $dq \wedge dp$ to $T_{(q,p)}({\rm Fix} \ \! G) \subset T_{(q,p)}(T^*\mathbb{R}^N)$ is nondegenerate. Let us first of all identify $T_{(q,p)}(T^*\mathbb{R}^N)$ by $T^*\mathbb{R}^N$ and $T_{(q,p)}({\rm Fix} \ \! G)$ by ${\rm Fix} \ \! G$ and moreover note that since $G$ is compact, it contains a unique left-invariant probability measure `$dP$', called the Haar-measure of $G$. We can therefore define the operator $${\rm av}_G:T^* \mathbb{R}^N \to {\rm Fix} \ \! G \ , \ v \mapsto \int_G P(v) dP$$ The operator ${\rm av}_G$ is a projection of $T^*\mathbb{R}^N$ onto ${\rm Fix} \ \! G$, called `averaging over $G$'. Now let $v \in {\rm Fix} \ \! G$ and $w \in T^*\mathbb{R}^N$. Then one easily computes that 
\begin{align} \nonumber &(dq \wedge dp)(v,{\rm av}_G(w))= (dq \wedge dp)(v,  \int_G P(w)d P) = \int_G (dq \wedge dp)(v,P(w))dP \\
&= \int_G (dq \wedge dp)(P(v),P(w))dP = \int_G (dq \wedge dp)(v,w)dP = (dq \wedge dp)(v,w)\nonumber \end{align}
where in the second equality we have used the linearity of $dq \wedge dp$ in its second argument, in the third equality the fact that $v\in{\rm Fix} \ \! G$, and in the fourth equality that every $P \in G$ is symplectic. We now observe that when $v \in {\rm Fix} \ \! G$ and $(dq \wedge dp)(v,w)=0$ for every $w \in {\rm Fix} \ G$, then $(dq \wedge dp)(v,w)=0$ even for every $w \in T^*\mathbb{R}^N$.  Hence $dq \wedge dp$, when restricted to ${\rm Fix} \ \! G \cong T_{(q,p)}({\rm Fix} \ \! G)$, is a nondegenerate anti-symmetric bilinear form. In other words, ${\rm Fix} \ \! G$ is a symplectic subspace of $T^*\mathbb{R}^N$. The final statement of this proposition follows trivially from this result.
\hfill $\square$ \\ \mbox{} \\
\noindent  Let us now look at a periodic FPU lattice with an even number $N=2n+2$ of particles. Then an invariant subsystem is formed by the fixed point set of the compact group $\langle S \rangle = \{{\rm Id}, S\}$:
$$\mbox{Fix}\ \! \langle S \rangle = \{(q,p)\in T^*\mathbb{R}^{N} | q_j = -q_{2n+2-j}, p_j = -p_{2n+2-j}\ \forall j\}$$
Clearly, in $\mbox{Fix}\ \! \langle S \rangle$, $q_0=q_{n+1}=p_0=p_{n+1}=0$. Thus we see that ${\rm Fix} \ \! \langle S \rangle$ is filled with solutions $(q_1(t), \ldots, q_N(t); p_1(t), \ldots, p_N(t))$ for which the  $(q_1(t), \ldots, q_n(t);$ $p_1(t),$ $\ldots, p_n(t))$ constitute the general solution curves of the FPU lattice with fixed endpoints and $n$ moving particles. Hence, the FPU lattice with fixed endpoints and $n$ particles is embedded in the  periodic lattice with $2n+2$ particles. By Proposition \ref{compactinvariant}, it can be described as a Hamiltonian system on ${\rm Fix} \ \! \langle S \rangle$, which has the restriction of $dq \wedge dp$ as symplectic form, and is determined by the Hamiltonian function $H|_{{\rm Fix} \ \!  \langle S \rangle}$.
As coordinates on ${\rm Fix} \ \! S$ one could choose $(q_1, \ldots, q_n; p_1, \ldots, p_n)$.

\section{Quasi-particles}
Of course, the representation of $D_N$ on $T^*\mathbb{R}^N$ is the sum of irreducible representations. It is quite natural to choose coordinates on $T^*\mathbb{R}^N$ that are adapted to these irreducible representations. For the periodic lattice, we thus make the following real-valued Fourier transformation.  For $1 \leq k < \frac{N}{2}$ define:
\begin{align} \nonumber 
Q_k \! = \! \sqrt{\frac{2}{N}} \!\! \sum_{j \in \mathbb{Z}/_{N \mathbb{Z}}} \hspace{-.2cm}  \sin(\frac{2jk\pi}{N}) q_j \ ,& \  
P_k \! = \! \sqrt{\frac{2}{N}}  \!\! \sum_{j \in \mathbb{Z}/_{N \mathbb{Z}}} \hspace{-.2cm} \sin(\frac{2jk\pi}{N}) p_j   \\ \nonumber
Q_{N-k} \! = \! \sqrt{\frac{2}{N}} \!\!  \sum_{j \in \mathbb{Z}/_{N \mathbb{Z}}} \hspace{-.2cm}  \cos(\frac{2jk\pi}{N}) q_j \ ,& \  
P_{N-k} \! = \! \sqrt{\frac{2}{N}}  \!\! \sum_{j \in \mathbb{Z}/_{N \mathbb{Z}}} \hspace{-.2cm} \cos(\frac{2jk\pi}{N}) p_j  \\ \nonumber
Q_{N} = \frac{1}{\sqrt{N}} \!\! \sum_{j \in \mathbb{Z}/_{N \mathbb{Z}}} \hspace{-.2cm} q_j   \ ,& \ P_{N} = \frac{1}{\sqrt{N}} \!\!  \sum_{j \in \mathbb{Z}/_{N \mathbb{Z}}} \hspace{-.2cm} p_j
\end{align}
and if $N$ is even:
$$Q_{\frac{N}{2}} = \frac{1}{\sqrt{N}}  \!\! \sum_{j \in \mathbb{Z}/_{N \mathbb{Z}}} \hspace{-.2cm} (-1)^j q_j   \  , \ P_{\frac{N}{2}} = \frac{1}{\sqrt{N}} \!\!  \sum_{j \in \mathbb{Z}/_{N \mathbb{Z}}} \hspace{-.2cm} (-1)^j p_j $$


\noindent The new coordinates $(Q,P)$ are called {\it quasi-particles} or {\it phonons}. The transformation $(q,p) \mapsto (Q,P), T^*\mathbb{R}^N \to T^*\mathbb{R}^N$ is symplectic and one can express the Hamiltonian in terms of $Q$ and $P$. If we write for (\ref{hamfpu}) $$H = H_2 +  H_3 +  H_4$$ where $H_2$ is a quadratic polynomial in $(q,p)$ and $H_3$ and $H_4$ cubic and quartic  polynomials in $q$, then we
 find that (see \cite{jackson}, \cite{poggiruffo} or \cite{Rink})
\begin{equation}
 \label{H2inphonon} H_2 =  \sum_{k = 1}^{N}\frac{1}{2}(P_k^2 + \omega_k^2 Q_k^2)
\end{equation}
in which for $k=1, \ldots, N$ the numbers $\omega_k$ are the well-known normal mode frequencies of the periodic FPU lattice: $$\omega_k := 2 \sin (\frac{k \pi}{N})$$
\noindent This means that written down in quasi-particles, the equations of motion of the harmonic lattice $(\alpha = \beta = 0)$  are simply the equations for $N-1$ uncoupled harmonic oscillators and, as $\omega_N=0$, one free particle. In fact, the Hamiltonian system is Liouville integrable in this situation. Integrals are for instance the linear energies
$$E_k := \frac{1}{2}(P_k^2 + \omega_k^2 Q_k^2) $$
\noindent The FPU model is of course much more interesting when the forces between the particles are nonlinear, i.e. when $\alpha$ or $\beta$ is nonzero. The normal modes then interact in a complicated manner that is governed by the Hamiltonians $H_r$ $(r=3,4)$, which are of the  form
\begin{equation} \label{Hrinphonon}
H_r = \sum_{\theta:|\theta|=r} c_{\theta} \prod_{k = 1}^{N-1} Q_k^{\theta_k}
\end{equation}
Here the $\theta$ are multi-indices and the $c_{\theta}$  are real coefficients.  Note that for every value of $\alpha$ and $\beta$, $H$ is  independent of $Q_N=\frac{1}{\sqrt{N}}\sum_j q_j$. Hence the total momentum $P_N= \frac{1}{\sqrt{N}}\sum_j p_j$ is a constant of motion and the equations for the remaining variables are completely independent of $(Q_N,P_N)$. It is common to set the latter coordinates equal to zero, thus remaining  with a system on $T^*\mathbb{R}^{N-1}$ with coordinates $(Q_1, \ldots, Q_{N-1}, P_1, \ldots, P_{N-1})$.  As $\omega_1, \ldots, \omega_{N-1}>0$, we can conclude by the Morse-Lemma or Dirichlet's theorem \cite{A&M}, that the origin $(Q,P)=0$ is a dynamically stable equilibrium of this reduced system.\\ \mbox{} \\
\noindent Assume again that $N=2n+2$. From the definition of the quasi-particles and the definition of $S$, we conclude that $S$ acts as follows in Fourier coordinates
\begin{align}\nonumber
S : \ (Q_1, \ldots, Q_{N-1};& P_1, \ldots, P_{N-1}) \mapsto \\
\nonumber 
(Q_1, \ldots, Q_n, -Q_{n+1}, \ldots, -Q_{N-1};& P_1, \ldots, P_{n}, -P_{n+1}, \ldots, -P_{N-1})
\end{align}
So that 
$$\mbox{Fix}\langle S \rangle = \{(Q,P)\in T^*\mathbb{R}^{N-1} \ |  \ Q_k = P_k = 0 \ \forall \ n+1 \leq k \leq N-1 \ \}$$ 
which is a symplectic manifold isomorphic to $T^*\mathbb{R}^n$. Using the coordinates $(Q_1, \ldots,$ $Q_n;$ $P_1,\ldots,$ $P_n)$ on $\mbox{Fix} \langle S \rangle$, the restriction of the symplectic form $\sum_{j=1}^{N} dQ_j \wedge dP_j$ to ${\rm Fix}\ \! \langle S \rangle$ is simply  $\sum_{j=1}^n dQ_j \wedge dP_j$. By Proposition \ref{compactinvariant}, the Hamiltonian of the fixed endpoint lattice thus  is simply the restriction of the periodic FPU Hamiltonian (\ref{H2inphonon}, \ref{Hrinphonon}) to $\mbox{Fix} \langle S \rangle$, that is 
\begin{equation}\label{hamfixed} \nonumber
H|_{\mbox{Fix}\langle S \rangle} =  \sum_{k = 1}^{n}\frac{1}{2}(P_k^2 + \Omega_k^2 Q_k^2) +  H_3(Q_1,\ldots,Q_n,0, \ldots, 0) +  H_4(Q_1,\ldots, Q_n, 0, \ldots, 0)
\end{equation}
To distinguish we have used the notation $\Omega_k := \omega_k = 2\sin(\frac{k\pi}{2n+2})$ $(1\leq k \leq n)$ for the linear frequencies of the fixed endpoint lattice.

\section{The Birkhoff normal form}
Nishida's idea was to study the Hamiltonian of the fixed endpoint lattice using
Birkhoff 
normalisation, which  is a way of constructing a symplectic near-identity
transformation of the phase-space with the purpose of approximating the original Hamiltonian system by a simpler one. The study of this `Birkhoff normal form' can lead to important conclusions about the
original system. For $r\geq 2$, let $\mathcal{F}_r$ be the finite-dimensional space of homogeneous $r$-th degree polynomials in $(Q,P)$ on $T^*\mathbb{R}^{N-1}$ and let $\mathcal{F}:=\bigoplus_{r\geq 2} \mathcal{F}_r$.  With the Poisson bracket $\{\cdot, \cdot\}: \mathcal{F} \times \mathcal{F} \to \mathcal{F}$ defined by
$$\{F,G\} := \sum_{k=1}^{N-1} \frac{\partial F}{\partial q_k} \frac{\partial G}{\partial p_k} - \frac{\partial F}{\partial p_k} \frac{\partial G}{\partial q_k} $$
$\mathcal{F}$ is a so-called graded Lie-algebra, which means that $\{\mathcal{F}_r, \mathcal{F}_s\} \subset \mathcal{F}_{r+s-2}$. With this definition, we have for each $F \in \mathcal{F}$, the `adjoint' linear operator $${\rm ad}_F: \mathcal{F} \to \mathcal{F} \ , \ G \mapsto \{F,G\}$$ We recall the following result, a complete proof of which can be found for instance in \cite{churchillkummerrod}, \cite{Cushman} and \cite{gaeta}. 

\begin{theorem}[Birkhoff normal form theorem] \label{Birkhoff} Let $H = H_2 + H_3 + \ldots \in \mathcal{F}$ be a Hamiltonian on $T^*\mathbb{R}^{N-1}$ such that $H_r \in \mathcal{F}_r$ for each $r$ and $$\mbox{ad}_{H_2}: G \mapsto \{H_2, G \} \ , \  \mathcal{F}_r \to \mathcal{F}_r$$ is semi-simple (i.e. complex-diagonalizable) for every $r$. Then for every finite $s \geq 3$ there is an open neighbourhood $0 \in U\subset T^*\mathbb{R}^{N-1}$ and a symplectic diffeomorphism $\Phi:U \to T^*\mathbb{R}^{N-1}$ with the properties that $\Phi(0)=0$, $D\Phi(0) = \rm{Id}$ and 
$$\Phi^*H=  H_2 + \overline{H}_3 + \ldots + \overline{H}_s + \mathcal{O}(||(Q,P)||^{s+1}) $$
where $${\rm ad}_{H_2}(\overline{H}_r) = 0 $$ for every $3\leq r \leq s$. The transformed and truncated Hamiltonian $\overline{H}:=H_2 + \overline{H}_3 + \ldots + \overline{H}_s$ is called a Birkhoff normal form of $H$ of order $s$.
\end{theorem}
{\bf Idea of proof:} For $H, F \in \mathcal{F}$, the curve $t \mapsto (e^{tX_{F}})^*H = H \circ e^{tX_F}$ in $\mathcal{F}$ satisfies the linear differential equation and initial condition $$\frac{d}{dt} (e^{tX_F})^*H = -{\rm ad}_{F} ((e^{tX_F})^*H ) \ , \ (e^{0X_F})^*H=H$$
This implies that $$(e^{-X_F})^*H  =  e^{{\rm ad}_F}(H) = H + \{F,H\} + \frac{1}{2}\{F, \{F, H\}\} + \ldots$$
The transformation $\Phi$ is now constructed as the composition of a sequence of time-$-1$ flows $e^{-X_{F_r}}$ $(3 \leq r \leq s)$ of Hamiltonian vector fields $X_{F_{r}}$ with $F_r \in  \mathcal{F}_r$.  The idea is that we first choose $F_3 \in \mathcal{F}_3$, such that $H$ is transformed into $$(e^{-X_{F_3}})^*H = \underbrace{H_2}_{\in \mathcal{F}_2} + \underbrace{H_3 + \{F_3, H_2\}}_{\in \mathcal{F}_3} + \underbrace{\ldots}_{\in \mathcal{F}_4 \oplus \mathcal{F}_5 \oplus \ldots}$$
When ${\rm ad}_{H_2}$ is semi-simple, then $\mathcal{F}_3 = {\rm ker \ ad}_{H_2} \oplus {\rm im \ ad}_{H_2}$ and we can decompose $H_3 = H_3^{ker} + H_3^{im}$ for uniquely determined $H_3^{ker} \in {\rm ker \ ad}_{H_2}$ and $H_3^{im} \in {\rm im \ ad}_{H_2}$. If we now choose $F_3$ such that ${\rm ad}_{H_2}(F_3) = H_3^{im}$, which obviously is possible, then $(e^{-X_{F_3}})^*H = H_2 + \overline{H}_3 + \ldots$ for $\overline{H}_3 = H_3 + \{F_3, H_2\} = H_3 -{\rm ad}_{H_2}(F_3) = H_3 - H_3^{im} = H_3^{ker} \in {\rm ker \ ad}_{H_2} $, i.e. ${\rm ad}_{H_2}(\overline{H}_3) = 0$. We continue by choosing $F_4 \in \mathcal{F}_4$ such that $(e^{-X_{F_4}})^*(( e^{-X_{F_3}})^*H) = H_2 + \overline{H}_3 + \overline{H}_4 + \ldots $ for which  ${\rm ad}_{H_2}(\overline{H}_4) = 0 $, etc. After $s-2$ steps we obtain $\overline{H}$ with the desired properties. \hfill $\square$  \\ \mbox{} \\
\noindent The normal form $\overline{H}$ is usually simpler than the original $H$ because it Poisson commutes with the quadratic Hamiltonian $H_2$. This firstly means that $H_2$ is a constant of motion for $\overline{H}$ and secondly that the flow $t \mapsto e^{tX_{H_2}}$ is a continuous symmetry of $\overline{H}$. \\
\indent Also, $H$ and $\overline{H}$ are symplectically equivalent modulo a small perturbation of order $\mathcal{O}(||(Q, P)||^{s+1})$. Studying $\overline{H}$ instead of $H$ thus means neglecting this perturbation term. So we make an approximation error, but this error is very small in the low energy domain, that is for small $||(Q, P)||$. With Gronwall's lemma, precise error estimates can be made. \\
\indent Finally, I would like to mention the ill-known bijective correspondence between the relative equilibria of the Birkhoff normal form and the bifurcation equations for periodic solutions obtained by Lyapunov-Schmidt reduction, as is explained in \cite{DuisMontecatini}.\\
\indent For Hamiltonian systems with symmetry, the following elegant and well-known result is often useful, see \cite{churchillkummerrod} and \cite{gaeta}:
\begin{theorem}\label{symnormform} Let $H = H_2 + H_3 + \ldots \in \mathcal{F}$ and $G$ be a group of linear symplectic symmetries of $H$. Then a normal form $\overline{H} = H_2 + \overline{H}_3 + \ldots + \overline{H}_s $ for $H$ can be constructed such that also $\overline{H}$ is $G$-symmetric.
\end{theorem}
This is obvious when one realizes that the `generating functions' $F_r$ of the proof of Theorem \ref{Birkhoff} can be chosen $G$-symmetric as well. \\ \indent  We shall also use the following result on normal forms of symmetric subsystems, which trivially follows from Proposition \ref{compactinvariant} and the proof of Theorem \ref{symnormform}, as the transformations $e^{-X_{F_r}}$ induced by symmetric Hamiltonian functions $F_r$  leave $\mbox{Fix}\ G$ invariant.
\begin{corollary} \label{beperking}
Let $H$ be a Hamiltonian function with compact symmetry group $G$ consisting of linear symplectic mappings. Then the normal form of $H|_{{\rm Fix}\ \!G}$ is simply the restriction of the symmetric normal form $\overline{H}$ of $H$ to ${{\rm Fix}\ \!G}$, i.e. 
$$\overline{H|_{{\rm Fix}\ \!G}} = \overline{H}|_{{\rm Fix}\ \! G}$$ 
\end{corollary} 
\mbox{} \\
This corollary tells us that it is sufficient to compute the normal form of the full system to know the normal forms of its symmetric subsystems. In particular, to find the normal form of an FPU lattice with fixed endpoints, it suffices to know the normal form of the appropriate periodic lattice. Normal forms of periodic lattices have been studied elaborately in \cite{Rink2}.

\section{Nishida's conjecture}\label{conjecture}
In his 1971 paper, Nishida proved the following result:
\begin{theorem}[Proven by Nishida in \cite{Nishida}]\label{GroteconjectureNishida} Consider the FPU lattice with fixed endpoints, $\alpha =0$, $\beta\neq 0$ and $n$ arbitrary. Assume moreover the fourth order nonresonance condition on the $\Omega_k = 2 \sin(\frac{k \pi}{2n+2})$ $(1\leq k \leq n)$ requiring that 
$$\sum_{k=1}^n (l_k - m_k)\Omega_k \neq 0 \ \forall \ l, m \in \{0,1,2,\ldots\}^n \ \mbox{with}\  \sum_{k=1}^n |l_k| + |m_k| = 4\  \mbox{and}\  \sum_{k=1}^n |l_k - m_k| \neq 0$$ Then the quartic Birkhoff normal form $\overline{H} = H_2 +  \overline{H}_4$ of the lattice is a function of the action variables $a_k:= E_k/\Omega_k$ $(1\leq k \leq n)$ only and is therefore integrable. Moreover it satisfies the Kolmogorov nondegeneracy condition $$\det \frac{\partial^2 \overline{H}}{\partial a_{k}\partial a_{k'}} \neq 0$$ This implies that almost all low-energy solutions of the $\beta$-lattice with fixed endpoints
 are quasi-periodic and move on
invariant tori. More precisely, the relative Lebesgue measure of all
these tori lying inside the
small ball $\{ 0 \leq  H \leq \varepsilon \}$, goes to $1$ as
$\varepsilon$ goes to $0$.
\end{theorem}
As we shall see later, the numbers 
$$\sum_{k=1}^n (l_k - m_k)\Omega_k  \ , \  {\rm for} \  \sum_{k=1}^n |l_k| + |m_k| = 4 $$ 
are simply the eigenvalues of ${\rm ad}_{H_2}$ on $\mathcal{F}_4$.  Nishida's requirement that they be nonzero except in the trivial case that $l_k= m_k$ for all $k$ thus just means that the subspace ${\rm ker \ ad}_{H_2} \in \mathcal{F}_4$ in which $\overline{H}_4$ must lie is very low-dimensional. It must therefore be remarked here that the integrability of the normal form follows almost trivially from Nishida's nonresonance assumption. Nishida's article consists mainly of the explicit computation of the normal form $\overline{H}$ of $H$ under the nonresonance assumption in order to check its nondegeneracy. \\ \indent But unfortunately, resonances do occur, implying that Nishida's nonresonance condition is often violated. We have for instance the relations
$$\sin(\pi/6)+ \sin(3\pi/14)-\sin(\pi/14)- \sin(5\pi/14)=0$$
$$\sin(\pi/6)+ \sin(13\pi/30)-\sin(7\pi/30)- \sin(3\pi/10)=0$$
$$\sin(\pi/2)+ \sin(\pi/10)-\sin(\pi/6)- \sin(3\pi/10)=0$$
which lead to a violation of Nishida's nonresonance condition if $n+1$ is a multiple of $21$ or $15$. \\ 
\indent Nishida refers to an unpublished result of Izumi proving a much stronger nonresonance condition on the $\Omega_k$ in special cases. The result states that no $\mathbb{Z}$-linear relations between the $\Omega_k$ exist if $n+1$ is a prime number or a power of $2$. I was not able to trace back Izumi's proof of this statement, but note that a more general result had already been obtained in 1959 by Hemmer \cite{Hemmer}, who actually derived an expression for the total number of independent $\mathbb{Z}$-linear relations between the $\Omega_k$ $(1\leq k \leq n)$ in terms of Euler's phi-function. It turns out that no $\mathbb{Z}$-linear relations exist if and only if $n+1$ is a prime number or a power of $2$. \\ \indent Moreover, as the above examples illustrate, resonance relations between $4$ eigenvalues exist for several $n$ and Nishida's condition is therefore sometimes violated. In this paper we will nevertheless prove `Nishida's conjecture' that his theorem holds without having to impose any nonresonance condition.

\section{Near-integrability}
Let us start with a review of some observation in \cite{Rink2} for the periodic FPU lattice. First of all we note that, as the symmetry $R$ is symplectic, $$(R^* \circ {\rm ad}_{H_2})(G) = R^*\{H_2, G\} = \{R^* H_2, R^* G\}= \{H_2, R^*G\} = ({\rm ad}_{H_2} \circ R^*)(G)$$
where we have used that $H_2$ is $R$-symmetric. From this result we read off that $R^*$ and ${\rm ad}_{H_2}$ commute as linear operators $\mathcal{F}_r \to \mathcal{F}_r$. This means that they can be diagonalized simultaneously. In \cite{Rink2} this is done by introducing new canonical coordinates $(Q,P) \mapsto (z, \zeta)$ as follows. For $1 \leq k <
\frac{N}{2}$, we define:  
\begin{align} \nonumber 
z_k &= \frac{1}{2}(P_{N-k} - iP_{k})  + \frac{
\omega_k}{2} (Q_k + i Q_{N-k})  \\ \nonumber 
z_{N-k} &= -\frac{1}{2}(P_{N-k}  -i P_{k}) + \frac{ \omega_k}{2} (Q_k + i Q_{N-k} ) \\ \nonumber 
\zeta_k &=  \frac{1}{2  \omega_k} (P_k - i P_{N-k} )
-\frac{1}{2}(Q_{N-k} +iQ_{k}) \\\nonumber
\zeta_{N-k} &= \frac{1}{2 \omega_k}(P_k - i P_{N-k} ) +
\frac{1}{2}( Q_{N-k}+ i Q_{k})  \nonumber
\end{align}
\noindent and if $N$ is even:
\begin{align}
z_{\frac{N}{2}} = \frac{1}{\sqrt{2}} (Q_{\frac{N}{2}} - \frac{i}{2}P_{\frac{N}{2}}) \ \ , \ \ 
&\zeta_{\frac{N}{2}} =  \frac{1}{\sqrt{2}}(P_{\frac{N}{2}} - 2 i
 Q_{\frac{N}{2}})  \nonumber
\end{align}
It is then not hard to compute that 
\begin{equation} \nonumber
H_2 = \sum_{1 \leq k < \frac{N}{2}} i \omega_k ( z_k \zeta_k - z_{N-k}
\zeta_{N-k} ) + i\omega_{\frac{N}{2}} z_{\frac{N}{2}}
\zeta_{\frac{N}{2}} 
\end{equation}
which implies that if $\Theta, \theta \in \{0, 1, 2, \ldots\}^{N-1}$ are
multi-indices, then
\begin{equation} \nonumber
\mbox{ad}_{H_2} :  \ z^{\Theta} \zeta^{\theta} \mapsto \nu(\Theta,
\theta) z^{\Theta} \zeta^{\theta} 
\end{equation}
 
\noindent in which $\nu$ is defined as

\begin{equation}\label{nu}
\nu(\Theta, \theta) := \sum_{1 \leq k < \frac{N}{2}} i\omega_k (
\theta_k - \theta_{N-k} - \Theta_k + \Theta_{N-k} ) \ +
i \omega_{\frac{N}{2}} ( \theta_{\frac{N}{2}} - \Theta_{\frac{N}{2}} ) 
\end{equation}
In other words, ${\rm ad}_{H_2}$ is diagonal with respect to the basis of $\mathcal{F}_r$ consisting of the monomials $z^{\Theta}\zeta^{\theta}$ for which $|\Theta| + |\theta|: = \sum_{j=1}^{N-1} |\Theta_j| + |\theta_j| = r$ and the corresponding eigenvalues are the $\nu(\Theta, \theta)$. In particular we observe that ${\rm ad}_{H_2}$ is semi-simple on every $\mathcal{F}_r$, so that Theorem \ref{Birkhoff} indeed applies. A $\mathbb{Z}$-linear relation in the frequencies $\omega_k$ is called a {\it resonance}.  For this reason, the monomials $z^{\Theta}\zeta^{\theta}$ for which $\nu(\Theta, \theta) = 0$ are called
resonant monomials. They are particularly important because they are exactly the ones that are not in ${\rm im \ ad}_{H_2}$ and thus, as is clear from the proof of Theorem \ref{Birkhoff}, the ones that
cannot be transformed away by Birkhoff normalisation. As $\Omega_k = \omega_k (1 \leq k \leq n)$, Nishida's nonresonance condition is a consequence of its analogon for periodic lattices, that can be formulated as follows: \begin{center}
{\it `When $|\Theta| + |\theta| = 4$ and $\nu(\Theta, \theta) = 0 $ then $\theta_{\frac{N}{2}} - \Theta_{\frac{N}{2}} = 0$  \\ and $\theta_k - \theta_{N-k} - \Theta_k + \Theta_{N-k}  = 0$ for each $1 \leq k <\frac{N}{2}$.'} \end{center}  
\noindent Of course, this condition is not valid either. \\
\indent On the other hand, one may compute, see \cite{Rink2}, that the operator $R^*: G \mapsto G \circ R$ acts as follows on the coordinate function $z_k, \zeta_k$:
\begin{align}\nonumber
R^*: \ \  z_k \mapsto  e^{\frac{2\pi i k}{N}} z_k,\ \ 
\zeta_k \mapsto  e^{-\frac{2\pi i k}{N}} \zeta_k, \ & \  z_{N-k} \mapsto
e^{\frac{2\pi i k}{N}} 
z_{N-k},\ \  \zeta_{N-k} \mapsto e^{-\frac{2\pi i k}{N}} \zeta_{N-k}\ , \\ \nonumber
z_{\frac{N}{2}} \mapsto - z_{\frac{N}{2}} \  \mbox{and} & \
\zeta_{\frac{N}{2}} \mapsto
-\zeta_{\frac{N}{2}} 
\end{align}

\noindent And as a result we conclude that, as promised, $R^*$ acts diagonally with respect to the monomials $z^{\Theta}\zeta^{\theta}$   as well: 
\begin{equation} \nonumber
R^*: \ z^{\Theta} \zeta^{\theta} \mapsto e^{\frac{2 \pi i \mu(\Theta,
\theta)}{N}} \ 
z^{\Theta} \zeta^{\theta} 
\end{equation}

\noindent in which  $\mu$ is defined as:

\begin{equation}\label{mu}
\mu(\Theta,\theta) := \sum_{1 \leq k < \frac{N}{2}} j (\Theta_k +
\Theta_{N-k} - \theta_k - \theta_{N-k} ) + \frac{N}{2} (
\Theta_{\frac{N}{2}} - \theta_{\frac{N}{2}} )  \ \ \mbox{mod} \ N 
\end{equation}
By Theorem \ref{symnormform} we now know that the normal form of the periodic FPU Hamiltonian must be a linear combination of monomials $z^{\Theta}\zeta^{\theta}$ that are both resonant and symmetric, i.e. for which $\nu(\Theta, \theta) = 0$ and $\mu(\Theta, \theta) = 0 \ \mod \ N$. The following theorem was proven in \cite{Rink2}. The proof below is considerably simpler though.
\begin{theorem}\label{hoofdstellingeigenwaarden} \hspace{1cm}\\
\\[-.3cm]
\noindent i) The set of multi-indices $(\Theta, \theta) \in
\mathbb{Z}_{\geq 0}^{N-1}$ for which $|\Theta| + |\theta| = 3, \ \mu(\Theta,
\theta) = 0\ \mbox{mod}\ N$ and $\nu(\Theta, \theta) = 0$ is empty.\\
\\[-.3cm]
\noindent ii) The set of
multi-indices $(\Theta, \theta) \in
\mathbb{Z}_{\geq 0}^{N-1}$ for which
$|\Theta| + |\theta| = 4, \ \mu(\Theta, \theta) = 0\ \mbox{mod} \ N$ and
$\nu(\Theta, \theta) = 0$ is contained in the set defined by the
relations $\theta_k - \theta_{N-k} - \Theta_k + \Theta_{N-k} =
\theta_{\frac{N}{2}} - \Theta_{\frac{N}{2}} = 0$. 
\end{theorem}
{\bf Proof:} \\
\\[-.3cm]
\noindent {\it i)} Suppose that $|\Theta|+|\theta| = 3$ and $\mu(\Theta, \theta)=0 \mod N$. Then we can conclude from looking closely at (\ref{mu}) and (\ref{nu}) that there must be integers $k, l, m \neq 0 \mod\ N$ with $k+l+m = 0 \mod \ N$ for which $\nu(\Theta, \theta) = 2i\sin(\frac{k\pi}{N}) +  2i\sin(\frac{l\pi}{N}) +  2i\sin(\frac{m\pi}{N}) =  2i\sin(\frac{k\pi}{N}) +  2i\sin(\frac{l\pi}{N}) -  2i\sin(\frac{k\pi}{N} + \frac{l\pi}{N}) $. Now I learnt the following trick from Frits Beukers: write $2 i \sin (\frac{k\pi}{N}) = x -1/x$ and $2 i \sin (\frac{l\pi}{N}) = y -1/y$ for some $x, y$ on the complex unit circle. Then $\nu(\Theta,\theta) = x-1/x + y- 1/y -xy + 1/xy = (1-x)(1-y)(1-xy)/xy$. This is zero only in the trivial cases that $x=1$ ($k=0\mod\ N$), $y=1$ ($l=0\mod\ N$) or $xy=1$ ($m=0\mod\ N$). But we already knew that $k,l,m\neq 0 \mod\ N$. The result also follows from the convexity of the sine function.\\ 
\\[-.3cm]
\noindent {\it ii)} The proof of the second statement is similar but more remarkable, and based on the fact that $2i\sin \alpha + 2i\sin \beta + 2i\sin \gamma - 2i\sin (\alpha + \beta + \gamma) = x-1/x + y- 1/y +z-1/z -xyz + 1/xyz = (1-xy)(1-xz)(1-yz)/xyz$, which again is zero in trivial cases only. $\hfill \square$ \\ \\
\noindent In spite of Theorem \ref{hoofdstellingeigenwaarden}, resonances do exist, as was illustrated by the examples in Section \ref{conjecture}. A full classification of third and fourth order resonance relations in the FPU eigenvalues is given in the Appendix to \cite{Rink2}. Resonance relations lead to several nontrivial resonant monomials. But according to Theorem \ref{hoofdstellingeigenwaarden} we now know that these nontrivial resonant monomials are not $R$-symmetric and hence cannot occur in the normal form of the periodic FPU lattice. As a first result, we immediately see now that there are no nonzero elements of $\mathcal{F}_3$ that are both resonant and $R$-symmetric. As a result, $\overline{H}_3 = 0$ automatically. \\
\indent To formulate a result for $\overline{H}_4$, we need to define the following {\it Hopf-variables}. For $1 \leq k < \frac{N}{2}$, let 
\begin{align}\label{abcd}\nonumber
a_k := \frac{1}{2 \omega_k}
(P_k^2 + P_{N-k}^2 + \omega_k^2 Q_k^2 + \omega_k^2 Q_{N-k}^2) \ &, \ 
b_k :=  Q_k P_{N-k} - Q_{N-k} P_k \\ 
c_k :=  \frac{1}{2 \omega_k}(P_{k}^2 - P_{N-k}^2 +
\omega_k^2 Q_{k}^2 - \omega_k^2 Q_{N-k}^2)  \ &, \  
d_k :=  \frac{1}{\omega_k} (P_k P_{N-k} + \omega_k^2 Q_k Q_{N-k}) \nonumber
\end{align}
and if $N$ is even
$$a_{\frac{N}{2}} := \frac{1}{2
\omega_{\frac{N}{2}}}(P_{\frac{N}{2}}^2 + \omega_{\frac{N}{2}}^2
Q_{\frac{N}{2}}^2)$$
\noindent Note that $H_2$ can  be expressed as 
\begin{equation} \nonumber
H_2 = \sum_{1 \leq k \leq \frac{N}{2}} \omega_k a_k 
\end{equation}
We moreover observe that when $N=2n+2$, the identities $a_{\frac{N}{2}} = b_k=c_k=d_k=0$ and $a_k = E_k/\Omega_k$ $(1 \leq k < \frac{N}{2})$ hold on ${\rm Fix} \ \! \langle S \rangle$, so that our definitions agree with the definition of $a_k$ in Theorem \ref{GroteconjectureNishida}. The following result was proven in \cite{Rink2} for the periodic FPU lattice. The proof consists of a careful analysis of the subspace of resonant and $\langle R,S\rangle$-symmetric polynomials in $\mathcal{F}_4$ with the help of Theorem \ref{hoofdstellingeigenwaarden}. It is not very deep and we do not repeat it here.
\begin{theorem}\label{eenvoud}
Let $H=H_2+ H_3+ H_4$ be the periodic FPU Hamiltonian (\ref{H2inphonon}, \ref{Hrinphonon}). Then there
is a unique quartic Birkhoff  normal form $\overline{H} = H_2 + \overline{H}_4$ of $H$ which is $\langle R, S \rangle$-symmetric. For this normal form we
have $\overline{H}_3 = 
0$, whereas $\overline{H}_4$ is a linear combination of the quartic terms $a_k a_{k'}$, $b_k b_{k'}$ $(1 \leq k , k' < \frac{N}{2})$ and if
$N$ is even also $a_{\frac{N}{2}} a_k$ $(1 \leq k \leq \frac{N}{2})$ and $d_k
d_{\frac{N}{2} - k} - c_k c_{\frac{N}{2}-k}$ $(1 \leq k \leq
\frac{n}{4})$. 
\end{theorem}

\begin{corollary}[Conjectured by Nishida in \cite{Nishida}]\label{corr0}  Independent of $n, \alpha$ and $\beta$, the quartic Birkhoff normal form $\overline{H}= H_2 + \overline{H}_4$ of the FPU lattice with fixed endpoints (\ref{hamfixed}) is integrable with integrals $E_k$. 
\end{corollary}
{\bf Proof:}  By Corollary \ref{beperking}, the Birkhoff normal form of (\ref{hamfixed}) is the restriction of the Birkhoff normal form of the periodic lattice with $N = 2n+2$ particles, to  $\mbox{Fix} \ \! \langle S\rangle$. But on $\mbox{Fix} \ \! \langle S\rangle$, we have that $b_k=c_k=d_k=0$ and $a_k =E_k/\Omega_k$. So according to Theorem \ref{eenvoud}, $\overline{H}_4$ is a quadratic function of the Poisson commuting $E_k$. So, clearly, is $H_2$. $\hfill \square$ \\ \\
\noindent Note that to prove the integrability of the normal form of the fixed endpoint lattice, we had to use the symmetry of the periodic lattice in which it is embedded.  It must also be remarked here that it is very exceptional for a high-dimensional resonant Hamiltonian system to have an integrable normal form. 
\\ \indent  Let us dwell a little longer on the dynamics of the normal form and consider the integral map $E:
T^*\mathbb{R}^{n} \to \mathbb{R}^{n}$ that sends $(Q, P)
\mapsto (E_1, \ldots, E_n)$. One checks that when $E_k>0$ for every $k$, the derivatives $DE_k(Q, P)$ are all linearly independent. As the level sets of $E$ are moreover compact,
the theorem of Liouville-Arnol'd ensures that for each $e = (e_1,\ldots,e_n)$ with $e_k>0$ for each $k$, the level set  $E^{-1}(\{e\})$ is a smooth $n$-dimensional torus.
\\
\indent To compute the flow on these tori, we transform to action-angle coordinates $(Q, P) \mapsto
(a,\varphi)$ as follows. Let $\mbox{arg}: \mathbb{R}^2
\backslash \{(0,0)\} \to \mathbb{R} / _{2 \pi \mathbb{Z}}$ be the argument
function, $\mbox{arg}: (r \cos \Phi, r \sin \Phi) \mapsto \Phi$ and define 
$$\varphi_k = \mbox{arg}(P_k, \Omega_k Q_k)\  , \  a_k = E_k/\Omega_k = \frac{1}{2\Omega_k}(P_k^2 + \Omega_k^2 Q_k^2) \ , \ 1 \leq k \leq n$$ 
With the formula $d\ \mbox{arg} (x,y) = \frac{x
dy - y dx}{x^2 + 
y^2}$, one can verify that $(\varphi, a)$ are canonical
coordinates: $ d Q \wedge dP = d\varphi \wedge da$. So in these coordinates the  equations of motion read $$\dot a_k =   0 \ ,\  \dot \varphi_k = \Omega_k + \frac{\partial \overline{H}_4(a)}{\partial
a_k} $$ 
This  simply defines periodic or quasi-periodic motion.  Remark: $(\phi, a)$ are sometimes called `symplectic polar coordinates'.

\section{Nondegeneracy}
To verify that the normal form $\overline{H}$ is nondegenerate, we
examine the frequency map $\Omega$ which assigns to each invariant torus
the frequencies of the flow on it:

\begin{equation} \nonumber
\Omega: a \ \mapsto \left( \Omega_1 + \frac{\partial
\overline{H}_4(a)}{\partial a_1}, \ldots, \Omega_{n} +
\frac{\partial \overline{H}_4(a)}{\partial a_{n}}\right) 
\end{equation}

\noindent The nondegeneracy condition of the KAM theorem requires that $\Omega$ be a local diffeomorphism, which is the case if and only if the constant derivative
matrix $\frac{\partial^2 \overline{H}_4}{\partial a_k \partial a_{k'}}$
is
invertible. To check this, we will unfortunately need to compute the Birkhoff normal form explicitly, where until now we had been able to avoid this.  In the next Theorem we shall present the normal form of the FPU
Hamiltonian in the case
that $H_3 = 0$, i.e. $\alpha = 0$. This lattice, that has no cubic terms, is usually
referred to as the $\beta$-lattice. 
\begin{theorem}[Conjectured by Nishida in \cite{Nishida}] \label{normaalvormbetastelling}  If $\alpha = 0$, then a quartic Birkhoff normal form of FPU lattice with fixed endpoints is given by $\overline{H}=H_2 + \overline{H}_4$, where
$$\overline{H}_4 = \frac{\beta}{2n+2}\left( \sum_{1\leq k<l\leq n} \frac{\Omega_k \Omega_l}{4}a_k a_l +\sum_{1\leq k \leq n} \frac{3\Omega_k^2}{32} a_k^2 \right)$$

\end{theorem}
{\bf `Proof':} The computation of the normal form had already been performed by Nishida \cite{Nishida} who obtained exactly the above normal form, but under the assumption that resonant monomials are absent in the lattice Hamiltonian. We now know that these monomials can not occur in the Hamiltonian as they are not $R$-symmetric in the corresponding periodic lattice. Hence Nishida's computation gave the correct answer. \\ \indent The reader can find  similar
computations in \cite{Rink2}, \cite{Rink} and \cite{Sanders} of the normal form of the $\beta$-lattice with periodic boundary conditions. We can therefore obtain the result also by substituting $a_k = E_k / \Omega_k$, $b_k = c_k = d_k = 0$ on ${\rm Fix} \ \! \langle S \rangle$ in the normal form of the periodic lattice that was obtained for instance in Theorem 10.1  in \cite{Rink2}.  $\hfill \square$ \\ \\
\noindent It is now an easy excercise to prove the invertibility of the matrix $\frac{\partial^2 \overline{H}_4}{\partial a_k \partial a_{k'}}$. Its nondegeneracy was also checked by Nishida himself by  applying elementary row and column operations to compute the determinant that turns out to be nonzero. Thus we conclude: 
\begin{corollary}[Conjectured by Nishida in \cite{Nishida}]  If $\alpha =0$ and $\beta\neq 0$, then the integrable quartic Birkhoff normal form $\overline{H} = H_2 +  \overline{H}_4$ of the FPU lattice with fixed endpoints (\ref{hamfixed}) satisfies the Kolmogorov nondegeneracy condition. Hence almost all low-energy solutions of the FPU lattice with fixed endpoints
 are quasi-periodic and move on
invariant tori. In fact, the relative measure of all
these tori lying inside the
small ball $\{ 0 \leq  H \leq \varepsilon \}$, goes to $1$ as
$\varepsilon$ goes to $0$.
\end{corollary}

\noindent  Nishida, and we, chose to compute normal form $H_2 + \overline{H}_4$ only for the $\beta$-lattice. This computation  is already quite long, but it becomes extremely hard when $\alpha \neq 0$. It should nevertheless also be
possible to write down an expression for the fixed endpoints normal form if $\alpha
\neq 0$. For checking Kolmogorov's condition 
this will actually be necessary. We know a priori that the resulting normal form will be integrable and depends quadratically on the $E_k$ (or $a_k$). It is very likely that for a large open set of $\alpha$ and $\beta$ the nondegeneracy condition holds and the KAM
theorem applies. Without computation
this is clear for $|\alpha| \ll |\beta|$ (and $n$ fixed) because then
the coefficients of the normal form can differ 
only slightly from those in Theorem \ref{normaalvormbetastelling}. 
\section{Acknowledgement}
The author would like to address many thanks to F. Verhulst, J.J. Duistermaat, G. Gallavotti and F. Beukers for many valuable remarks and comments, without which this paper would not have been written. 

\bibliography{/Users/brink/Documents/Preferences/Bibtex/bibliografie}
\bibliographystyle{amsplain}

\end{document}